\documentclass[11pt]{article}

\usepackage{epsfig}
\usepackage{epstopdf}  
\usepackage{amsmath}
\usepackage[normalem]{ulem}

\renewcommand\sout{\bgroup \color{blue} \ULdepth=-.5ex \ULset}
\usepackage{color}
\usepackage{latexsym}
\usepackage{amssymb}
\usepackage{dsfont}
\usepackage{multirow}
\usepackage{epsfig}

\newcommand{\bea}{\begin{eqnarray}}
\newcommand{\eea}{\end{eqnarray}}
\newcommand{\be}{\begin{equation}}
\newcommand{\ee}{\end{equation}}

\newcommand{\lqcd}{\Lambda_\mathrm{QCD}}
\def\dpdf#1{D_{#1}}

\newlength\savedwidth

\title{\bf{Double parton scattering: 
\\
a study of the effective cross section
\\
within a Light-Front quark model}}
\author{Matteo Rinaldi, Sergio Scopetta  \\
Dipartimento di Fisica e Geologia,
Universit\`a degli Studi di Perugia, 
\\ 
and Istituto Nazionale di Fisica Nucleare,
Sezione di Perugia, 
\\
via A. Pascoli, I - 06123 Perugia, Italy	
\\
Marco~Traini
\\
Dipartimento di Fisica, Universit\`a degli studi di Trento, and
INFN - TIFPA, \\
Via Sommarive 14, I - 38123 Povo (Trento), Italy
\\
Vicente~Vento \\
Departament de Fisica Te\`orica, Universitat de Val\`encia
\\
and Institut de Fisica Corpuscular, Consejo Superior de Investigaciones
\\
Cient\'{\i}ficas, 46100 Burjassot (Val\`encia), Spain
}

\begin{document}
\maketitle{}

\begin{abstract}

We present a calculation of the effective cross section $\sigma_{eff}$, 
an important ingredient
in the description of double parton scattering in proton-proton collisions. 
Our theoretical approach makes use of a Light-Front quark model as 
framework to calculate the double parton distribution functions at 
low-resolution scale. QCD evolution is implemented to reach the experimental 
scale. The obtained $\sigma_{eff}$, when averaged over the longitudinal  
momentum fractions of the interacting partons, $x_i$, 
is consistent with the present experimental scenario.
However the result of the complete calculation
shows a dependence of $\sigma_{eff}$ on $x_i$, a feature not 
easily seen in the available data, probably  because of their low accuracy. 
Measurements of $\sigma_{eff}$ in restricted $x_i$ regions are 
addressed to obtain indications on 
double parton correlations, a novel and interesting aspect of the 
three dimensional structure of the nucleon.

\end{abstract}




\newpage

\section{Introduction}

Multi Parton Interactions (MPI),  occurring when more than one 
parton scattering
takes place in the same hadron-hadron collision, have been discussed in 
the literature 
since long time ago \cite{paver} and are presently attracting considerable attention, thanks to 
the possibilities offered by the Large Hadron Collider (LHC) (see Refs. 
\cite{gaunt,diehl_1,manohar_1,bansal,ciurek} for recent reports).
In particular, the cross section for double parton scattering (DPS), the simplest MPI process,
depends on peculiar non-perturbative quantities, the double parton distribution functions (dPDFs),
describing the number density of two partons with given longitudinal momentum fractions 
and located at a given transverse separation in coordinate space.
dPDFs are naturally related to parton correlations and to the three-dimensional (3D) nucleon 
structure, as discussed also in the past \cite{calucci}.

No data are available for dPDFs and their calculation using non perturbative methods is cumbersome.
A few model calculations have been performed, to grasp the most relevant features 
of dPDFs \cite{manohar_2,noi1,noi2}. 
In particular, in Ref. \cite{noi2} a Light-Front (LF) Poincar\`e covariant approach,
able to reproduce the essential sum rules of dPDFs, has been described.
Although it has not yet been possible to extract dPDFs from data,
a signature of DPS has been observed and measured in several 
experiments \cite{afs,data0,data2,data3,data4,data5}:  
the so called  ``effective cross section'', $\sigma_{eff}$.
Despite of large errorbars, the present experimental scenario 
is consistent with the idea 
that $\sigma_{eff}$ is constant w.r.t. the center-of-mass energy 
of the collision.

In this letter we present a predictive study of  $\sigma_{eff}$ which makes use of the LF quark
 model  approach to dPDFs developed in Ref. \cite{noi2}.

The definition of  $\sigma_{eff}$ is reviewed in the next section where the present experimental situation is also summarized. Then  the results of our approach are presented critically discussing the dynamical dependence of  $\sigma_{eff}$ in view of future experiments. Conclusions are drawn in the last section.

\section{The effective cross section}

The effective cross section, $\sigma_{eff}$, is defined through the so called ``pocket formula'', which reads, if final states $A$ and $B$ are produced in a DPS process (see, e.g., \cite{bansal}):

\begin{eqnarray}
\label{pocket}
\sigma_{eff} = { m \over 2 }
{ \sigma_A^{pp'} \sigma_B^{pp'} \over 
\sigma^{pp}_{double}
}~.
\end{eqnarray}
$m$ is a process-dependent combinatorial factor: $m= 1$
if $A$ and $B$ are identical and $m=2$ if they are different. $\sigma_{A(B)}^{pp'}$ is the differential
cross section for the inclusive process $pp' \rightarrow A (B) + X$,  naturally defined as:

\begin{eqnarray}
\label{s_single}
\sigma_{A}^{pp'}(x_1,x_1',\mu_1) & = &
\sum_{i,k}  F_{i}^{p} (x_{1},\mu_{1})
F_{k}^{p'} (x_{1'},\mu_{1})\,
\hat \sigma_{ik}^{A}(x_{1},x_{1'},\mu_{1})~,  \\
\sigma_{B}^{pp'}(x_2,x_2',\mu_2) & = &
\sum_{j,l}  F_{j}^{p} (x_{2},\mu_{2})
F_{l}^{p'} (x_{2'},\mu_{2})\,
\hat \sigma_{jl}^{B}(x_{2},x_{2'},\mu_{2})~,
\end{eqnarray}
where $F_{i(j)}^p$ is a one-body parton distribution function
(PDF)
with $i,j,k,l = \{q, \bar q, g \}$, $\mu_{1(2)}$ is
the renormalization scale for the process $A (B)$,
$\sigma^{pp}_{double}$, the remaining ingredient in Eq. (\ref{pocket}),
appears in the natural definition of the cross section
for double parton scattering:

\begin{eqnarray}
\sigma_{d} = 
\int 
\sigma^{pp}_{double}(x_1,x_1',x_2,x_2',\mu_1,\mu_2)
\,d x_1 d x_1' d x_2 d x_2'~,
\label{3}
\end{eqnarray}
and reads:

\begin{eqnarray}
\sigma^{pp}_{double}(x_1,x_1',x_2,x_2',\mu_1,\mu_2)& = &
{m \over 2} 
\sum_{i,j,k,l} \int
D_{ij}(x_1,x_2; {\bf {k_\perp}},\mu_1,\mu_2)\,
\hat \sigma_{ik}^A(x_1,x_1')\,
\hat \sigma_{jl}^B(x_2,x_2')\,
\nonumber
\\
& \times &
D_{kl}(x_1',x_2'; {\bf {- k_\perp}},\mu_1,\mu_2) 
{d {\bf {k_\perp}} \over (2 \pi)^2}~.
\label{4}
\end{eqnarray}
In the above equation,
${\bf {k_\perp}}$ ($-{\bf {k_\perp}}$) is the transverse momentum
unbalance of the parton 1 (2),
conjugated to the relative distance ${\bf {r_\perp}}$ (the reader
should not confuse ${\bf {k_\perp}}$ with the intrinsic momentum
of the parton, argument of transverse momentum dependent parton 
distributions).
The quantity
$D_{ij}(x_1,x_2; {\bf {k_\perp}})$,
called sometimes $_2GPDS$ \cite{blok_1,blok_2}, is therefore the
Fourier transform
of the so called double distribution function,
$D_{ij}(x_1,x_2; {\bf {r_\perp}})$, which represents the
number density of partons pairs $i,j$ with longitudinal momentum 
fractions $x_1,x_2$ , respectively, at a transverse separation
${\bf {r_\perp}}$ in coordinate space.
dPDFs, describing soft Physics,
are nonperturbative quantities.

\noindent Two main assumptions are usually made for the evaluation of dPDFs:
\\
a) factorization of the transverse separation and the momentum fraction 
dependence:
\begin{equation} 
\label{app1}
\dpdf{ij}(x_1,x_2,{\bf {k_\perp}},\mu) = \dpdf{ij}(x_1,x_2,\mu) \,
T({\bf {k_\perp}},\mu)
\,;\end{equation}
b)
factorized form also for the $x_1,x_2$ dependence:

\begin{equation}
\label{app2}
\dpdf{ij}(x_1,x_2,\mu) 
 = F_{i}(x_1,\mu) \, F_{j}(x_2,\mu)\, \theta(1-x_1-x_2) (1-x_1-x_2)^n\,.
\end{equation}
The expression $\theta(1-x_1-x_2) (1-x_1-x_2)^n$,
where $n>0$ is a parameter to be fixed phenomenologically,
introduces the natural
kinematical constraint $x_1+x_2 \leq 1$ (in Eqs. (\ref{app1})
and (\ref{app2}) the same scale $\mu=\mu_1,\mu_2$ is assumed, for brevity). 

\noindent One comment about the physical meaning 
of $\sigma_{eff}$ is in order.
\begin{figure}[h]
\begin{center}
\vskip -1cm
\epsfig{file=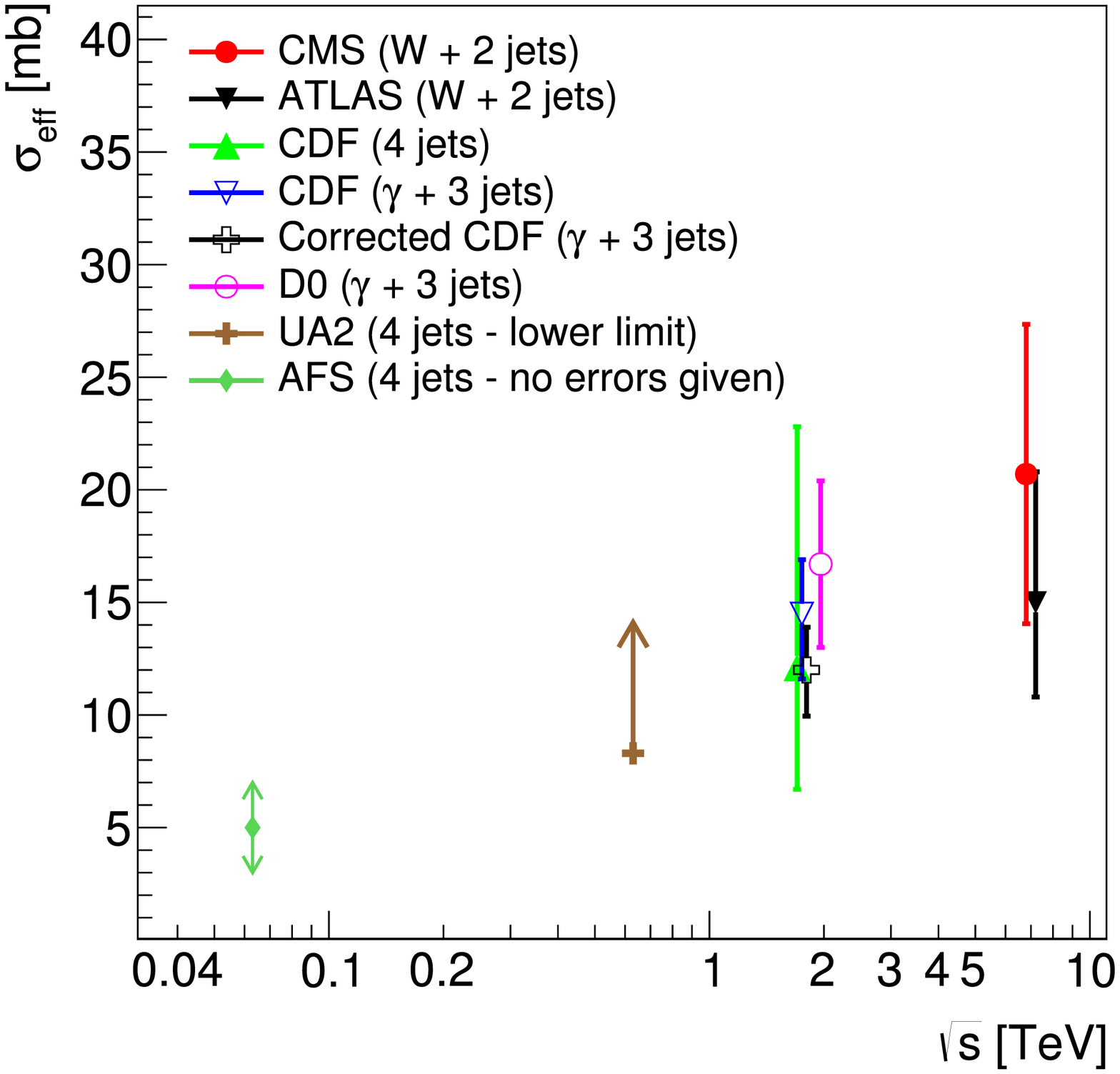, width=8cm}
\caption{Centre-of-mass energy dependence of $ \sigma_{eff}$
measured by different experiments using
different processes
\cite{afs,data0,data2,data3,data4,data5}.
The figure is taken from \cite{data5}}
\end{center}
\label{fig:Fig1}
\end{figure}
In Eq. (\ref{pocket}), if the occurrence of the process $B$ were not biased
somehow by that of the process $A$, instead of the ratio
$\sigma_B/\sigma_{eff}$ one would read
$\sigma_B/\sigma_{inel}$, representing the probability
to have the process $B$ once $A$ has taken place assuming rare
hard multiple collisions.
The difference between $\sigma_{eff}$ 
and $\sigma_{inel}$ 
measures therefore correlations between the interacting partons
in the colliding proton. 

Let us discuss now the dynamical dependence of $\sigma_{eff}$
on the fractional momenta $x_1,x_1',x_2,x_2'$.
By inserting Eqs. (2-5) in Eq. (1), and
omitting the dependence on the renormalization scales
for simplicity,
one gets the following expression for $\sigma_{eff}$:

\begin{eqnarray}
\label{final}
\sigma_{eff} (x_1,x_1',x_2,x_2') = 
{
\left\{ \sum_{i,k }  F_{i}^{p} (x_{1})
F_{k}^{p'} (x_{1'})\,
\hat \sigma_{ik}^{A}(x_{1},x_{1'}) \right\}
\left\{ \sum_{j,l }  F_{j}^{p} (x_{2})
F_{l}^{p'} (x_{2'})\,
\hat \sigma_{jl}^{B}(x_{2},x_{2'}) \right\}
\over
\sum_{i,j,k,l} 
\hat \sigma_{ik}^A(x_1,x_1')\,
\hat \sigma_{jl}^B(x_2,x_2')
\int
D_{ij}(x_1,x_2; {\bf {k_\perp}})\,
D_{kl}(x_1',x_2'; {\bf {-k_\perp}}) 
{d {\bf {k_\perp}} \over (2 \pi)^2}
}~.
\end{eqnarray}

Eq.(\ref{final}) clearly shows the dynamical origin of the
dependence of $\sigma_{eff}$
on the fractional momenta $x_1,x_1',x_2,x_2'$. Even within the ``zero rapidity region'', ($y=0$), where $x_1=x_1'$, $x_2=x_2'$, such a dependence, although simplified, is still effective.

Assuming that heavy flavors are not relevant in the process, the dependence on the ``parton type'', 
$i=q,\bar q,g$, of the elementary cross section is basically \cite{4}:

\begin{eqnarray}
\hat \sigma_{ij} (x,x') = C_{ij} \bar \sigma(x,x')~,
\label{cij}
\end{eqnarray}
where $\bar \sigma(x,x')$ is a universal function, and $C_{ij}$
are color factors which stay in the ratio:

\begin{eqnarray}
\label{color}
C_{gg}:C_{qg}:C_{qq}=1:(4/9):(4/9)^2~.
\end{eqnarray}
Using  Eq. (\ref{cij}), Eq. (\ref{final}) simplifies considerably:

\begin{eqnarray}
\label{simple}
\sigma_{eff} (x_1,x_1',x_2,x_2') = 
{
\sum_{i,k,j,l }  F_{i} (x_{1})
F_{k} (x_{1}')
F_{j} (x_{2})
F_{l} (x_{2}')
C_{ik}
C_{jl}
\over
\sum_{i,j,k,l} 
C_{ik}
C_{jl}
\int
D_{ij}(x_1,x_2; {\bf {k_\perp}})
D_{kl}(x_1',x_2'; {\bf {-k_\perp}}) 
{d {\bf {k_\perp}} \over (2 \pi)^2}
}
~.
\end{eqnarray}

The present experimental scenario is illustrated in Fig.1. The experiments 
\cite{afs,data0,data2,data3,data4,data5}, at different values 
of the center-of-mass
energy,$\sqrt{s}$, and with different final states,  
explore different regions of $x_i$.
Experiments at high $\sqrt{s}$ access low $x_i$ regions, in general.
The old AFS data \cite{afs} are in the valence region 
($0.2 \leq x_i \leq 0.3$),
the Tevatron data \cite{data2, data3} 
are in the range $0.01 \leq x_i \leq 0.4$ while
the recent LHC data \cite{data4,data5} cover a lower average $x_i$ range
and are dominated by the glue distribution. 

Remarkably the experimental evidences are compatible with a constant value of  
$\sigma_{eff}$ in Eq.(\ref{pocket}), the $x_i$-dependence 
being probably hidden 
within the experimental uncertainties. 
In fact one should stress that the knowledge of the $x_i$-dependence 
of $\sigma_{eff}$ 
would open the access to information on the $x_i$-dependence  of the 
dPDFs $D_{ij}(x_1,x_2; {\bf {r_\perp}})$, 
entering the definition of $\sigma_{eff}$: a direct 
way to access the 3D nucleon structure \cite{calucci}. 
Nowadays, the aspects of the
3D nucleon structure related to the transverse position
of partons are investigated through hard-exclusive 
electromagnetic processes, such as deeply virtual Compton scattering
(DVCS), extracting the Generalized Parton Distributions (GPDs)
(see Ref. \cite{gui} for recent results).
The information encoded in DPS, dPDFs and in $\sigma_{eff}$,
in its full $x_i$ dependence, are anyway different and complementary
to those provided by GPDs in impact parameter space.
While the latter quantities are one-body densities, depending on the distance, of the interacting 
parton with given $x$, from the transverse center of the target, in DPS one is sensitive to 
the {\it relative} distance between two partons with  given longitudinal momentum fractions. 
In other words, the investigation of dPDFs from DPS,  is relevant to know, for example, the  
average transverse distance of two fast partons or two slow partons: 
a very interesting dynamical feature, not accessible through GPDs.

%
%

\section{{\bf Light-Front quark model calculation of the effective cross section}}

dPDFs have a non-perturbative nature, and, at present, cannot be calculated  in  QCD. 
However they can be explicitly calculated, at a low resolution scale, $Q_0 \sim \lqcd$, using quark models, as extensively done for the usual PDFs.  
The results of these calculations should be then evolved using perturbative QCD (pQCD)
in order to match data taken at a momentum scale $Q>Q_0$. The procedure is nowadays 
well established (see, e.g., Ref. \cite{trvlc} and references therein).
The QCD evolution procedure of dPDFs (from  $Q_0$ to $Q>Q_0$) is  
well known~\cite{Kirschner,Shelest:1982dg},  and currently implemented in 
a systematic way 
(see Ref. \cite{diehl_1,ruiz}
and references therein).

The first model calculations of dPDFs in the valence region, at the hadronic scale $Q_0$, have been presented in a bag model framework ~\cite{manohar_2}, 
and in a constituent quark model (CQM), ~\cite{noi1}. Of course CQM have the 
peculiar advantage of including correlations in a way 
consistent with the quark dynamics, 
from the very beginning, a property that the bag model cannot fulfill. 


In particular the fully Poincar\'e covariant Light-Front model approach we developed in 
Ref. \cite{noi2} respects relevant symmetries, broken in the descriptions of 
Refs.\cite{manohar_2,noi1},  allowing for a correct evaluation of the Mellin 
moments
of the distributions and, consequently, 
for a precise pQCD evolution to high momentum transfer.
In this way our model calculations can be relevant for the analysis of high-energy data.



The model, extensively  applied to the evaluation  of different parton distributions, 
(see, e.g., Refs.
\cite{LF2,pasquini,marco2} and references therein),
is a good candidate to grasp the most relevant features of dPDFs.
(see Ref. \cite{LF2} for details). 
For the present study it is enough to recall that the proton state is given by a spatial wave function
and an SU(6) symmetric spin-isospin part.
The spatial part is numerical solution of a relativistic Mass equation, dynamically
responsible for the presence of  correlations between the two quarks in the CQM wave function
(a non-relativistic version of the model was introduced in Ref. \cite{santop}).
The Light-Front calculations of $\dpdf{ij}(x_1,x_2,{\bf k_\perp},\mu)$, 
in Ref.~\cite{noi2}, shows
that the factorization of Eq. (\ref{app1}) is basically valid, 
but the common assumption of Eq. (\ref{app2}) is strongly violated.  Besides, the strong
correlation effects present at the scale of the model are still sizable, in the valence region, at the experimental scale, i.e. after QCD evolution.
At the low values of $x$, presently studied at the LHC, 
the correlations become less relevant, 
although their effects are still important for the spin-dependent contributions 
to unpolarized proton scattering.

We have explicitly calculated single and double parton distributions 
entering Eq.(\ref{simple}), and then $\sigma_{eff}$
relying on the natural assumption Eq. (\ref{color}) only. 
We adhere, in addition, to the common choice of a single 
renormalization scale $\mu_1 = \mu_2 = \mu_0$, where $\mu_0$ 
has to be interpreted, in the present approach, as 
the hadronic scale where only valence quarks $u$ and $d$ are present.  
Considering the symmetries of our model, one has $u(x,\mu_0) = 2 d(x,\mu_0)$,
$D_{u,u}(x_1,x_2, k_\perp,\mu_0)=2 D_{u,d}(x_1,x_2, k_\perp,\mu_0)$ and
Eq. (\ref{simple}) simplifies to

\begin{eqnarray}
\label{simplem}
\sigma_{eff} (x_1,x_1',x_2,x_2',\mu_0) = 
{81
u(x_{1},\mu_0)\,
u (x_{1}',\mu_0)\,
u (x_{2}.\mu_0)\,
u (x_{2}',\mu_0)
\over
64 \int
D_{uu}(x_1,x_2; {\bf {k_\perp}},\mu_0)\,
D_{uu}(x_1',x_2'; {\bf {-k_\perp}},\mu_0)\, 
{d {\bf {k_\perp}} \over (2 \pi)^2}
}
~.
\end{eqnarray}

Since the experimental data in the valence region
are mainly restricted at zero rapidity ($y=0$), 
where $x_i = x_i'$, one remains with 
\begin{eqnarray}
\label{simplest}
\sigma_{eff} (x_1,x_2,\mu_0) = 
{81
u(x_{1},\mu_0)^2
u (x_{2}.\mu_0)^2
\over
64 \int
D_{uu}(x_1,x_2; {\bf {k_\perp}},\mu_0)^2
{d {\bf {k_\perp}} \over (2 \pi)^2}
}~.
\end{eqnarray}

\begin{figure}[t]
\begin{center}
\epsfig{file=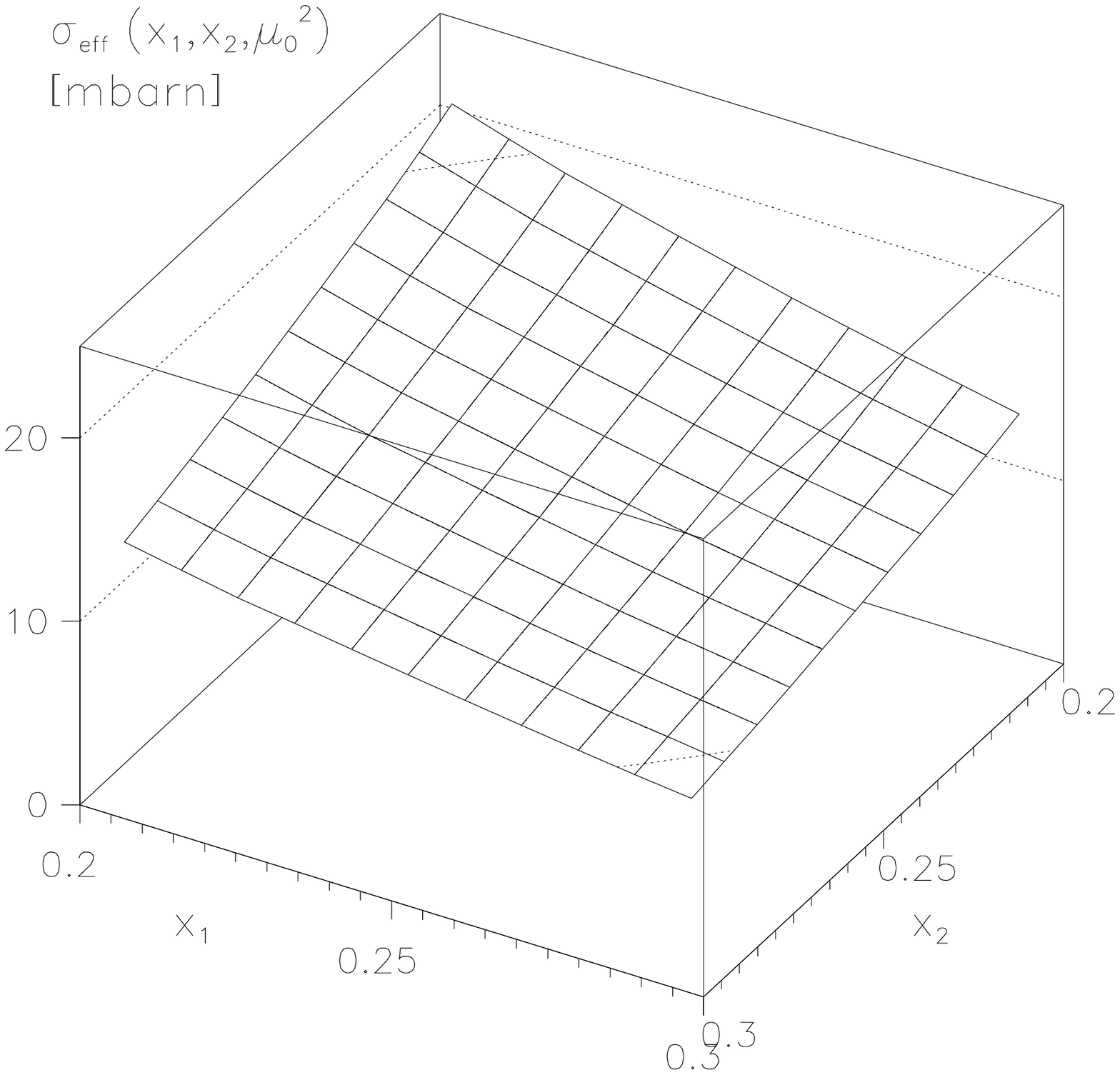,
width=6.cm
}
\epsfig{file=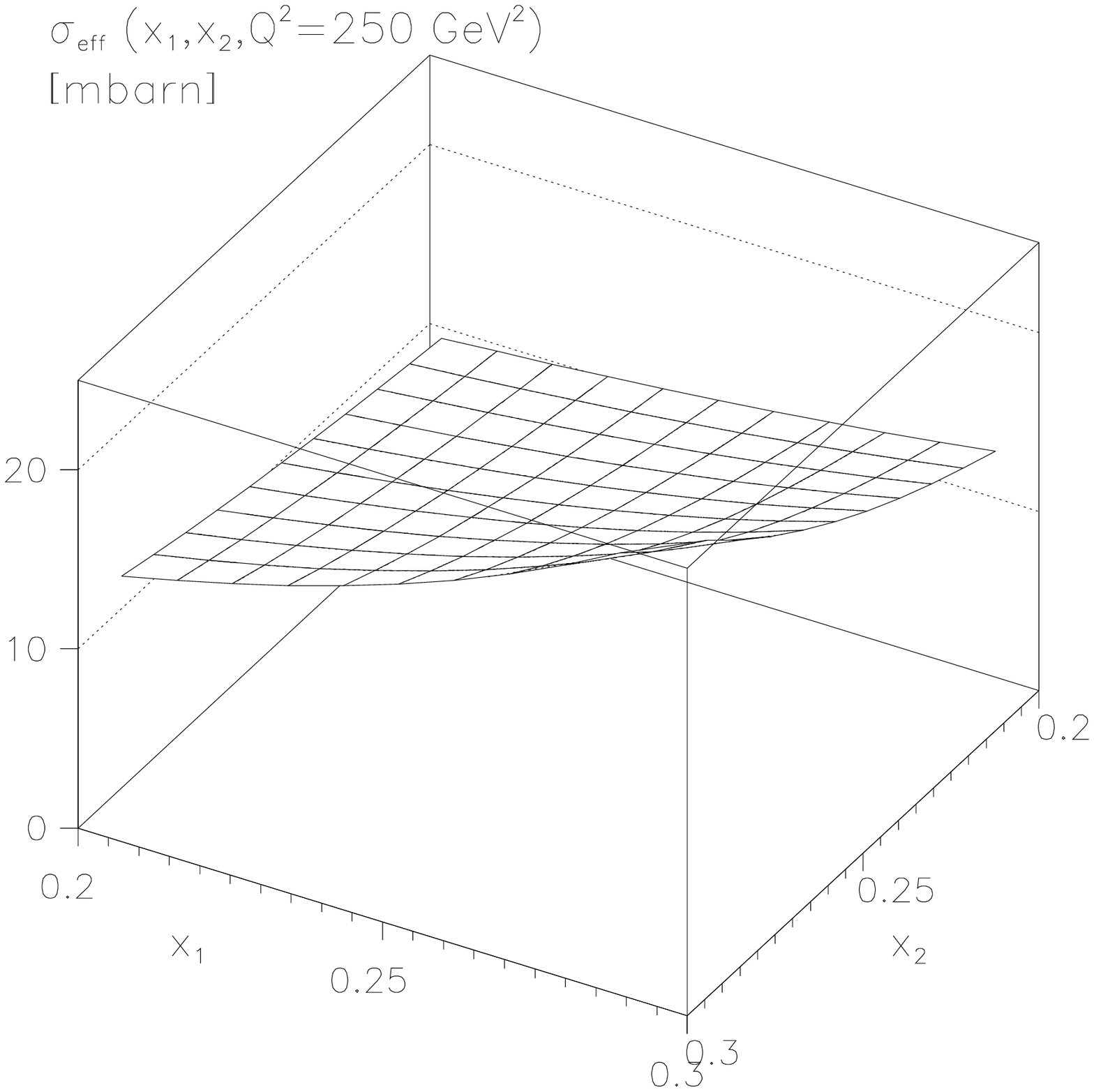, 
width=6.cm
}
\end{center}
\label{fig:Fig2}
\caption{$ \sigma_{eff}(x_1,x_2,Q^2)$ for the values of $x_1, x_2$
measured in Ref. \cite{afs}. Left panel:
hadronic scale; right panel: $Q^2$ = 250 GeV$^2$.}
\end{figure}

In order to illustrate our results we will concentrate on the valence region where the present model is more predictive. In particular we concentrate on the kinematics of the old AFS data \cite{afs}, which means $y=0$ ($x_1=x_1',x_2=x_2'$) and  $0.2 \leq x_{1,2} \leq 0.3$.
The average momentum scale, again assumed to be the same
for the processes initiated by the two different collisions,  turns out to be $Q^2 \simeq 250$ GeV$^2$.
The results of the calculations are shown in Fig. 2, at the scale of the model, $\mu_0^2 \simeq 0,1$ GeV$^2$, and after non-singlet evolution to $Q^2$ (details on the fixing of the  hadronic scale and on the calculation of the QCD evolution can be found in Ref. \cite{noi2}).

What is immediately seen is an $x_{1,2}$ dependence of the results, 
which change up to 100$\%$ 
even in this narrow kinematical range. Such a dependence is found at both the experimental and 
the model scale. The slope of the surface in the right panel of Fig. 2  is inverted w.r.t. that in the left panel.
It is not a surprising feature, due to the different evolution properties of the numerator and denominator in Eq. (\ref{simplem}) and consistent with the evolution calculated in Ref. \cite{noi2}.
This $x$ dependence, found at the hadronic scale as well as at high $Q^2$, can be attributed to different dynamical and kinematical properties:

\begin{enumerate}

\item both the numerator and  the denominator vanish quickly with $x_i$ through the valence region,
but the latter vanishes faster, mainly due to the  kinematic constraint $x_1+x_2 \leq 1$ of the dPDF, a quantity  appearing only in the denominator. The LF model correctly reproduce such a kinematical constrain;

\item the correlations introduced by the LF dynamics and effective both in the $x_i$ and $k_\perp$ dependence;

\item the correlations induced by the pQCD evolution in the valence region.

\end{enumerate}

In order to be more intuitive let us restrict to two different 
extreme scenarios:

\begin{itemize}

\item[i)] At very low-$x$  gluons are strongly dominating
(this is the hypothesis in \cite{blok_1}, partially corrected in \cite{blok_2}), 
so that it is enough to consider $i,j,k,l=g$. Assuming, in addition, 
a fully factorized approach: $D_{gg}(x,x',{\bf {k_\perp}}) = F_g(x)F_g(x')g({\bf {k_\perp}})$, 
$\sigma_{eff}$ becomes:

\begin{eqnarray}
\sigma_{eff} (x_1,x_1',x_2,x_2') \rightarrow  \sigma_{eff}=
{1 \over \int
g^2({\bf {k_\perp}})
{d {\bf {k_\perp}} \over (2 \pi)^2}
}~.
\label{eq:sigma_gg}
\end{eqnarray}

A similar assumption is used in Ref.  
\cite {blok_1} to obtain  an estimate of $\sigma_{eff}$ 
which turns out to be about twice the experimental value. 
Obviously, the validity of Eq.(\ref{eq:sigma_gg}) is spoiled by correlation 
effects and restricted to very low-$x_i$. 
The problems related to the uncorrelated ansatz are discussed in a 
number of papers
(see, e.g., Ref.~\cite{gaunt,snig1,cattaruzza,muld}). 
In particular, in the valence region this assumption
is not supported by model calculations \cite{manohar_2,noi1,noi2}
and it is certainly untrue in pQCD, being also spoiled by QCD evolution.
In other words, several arguments lead to the conclusion that,
in general, $\sigma_{eff}$ should be
$x_i$ dependent, namely: breaking of the
factorization ansatz; the QCD evolution; contribution
of more than one parton type (not only gluons as at very low
$x_i$) to the DPS cross section. 

\item[ii)]  Let us now consider a simple way to reduce the results of 
our calculation to 
a fully factorized approach to dPDFs, following the hypothesis 
often assumed (cfr. Eqs. (\ref{app1}), (\ref{app2})):

\begin{eqnarray}
\label{facto}
D_{uu}(x_1,x_2; {\bf {k_\perp}},\mu_0)
=
u(x_1,\mu_0) u(x_2, \mu_0) f_{uu} ({\bf {k_\perp}})~,
\end{eqnarray}

where a natural definition for the ``effective form factor'',
$f_{uu} ({\bf {k_\perp}})$, in our approach, is 

\begin{eqnarray}
\label{fuu}
f_{uu} ({\bf {k_\perp}}) ={1 \over 4}
\int d x_1 d x_2 D_{uu}(x_1,x_2; {\bf {k_\perp}},\mu_0)~,
\end{eqnarray}

a quantity which turns out to be scale independent. Within this approximation,
Eq. (\ref{simplem}) yields:

\begin{eqnarray}
\sigma_{eff} (x_1,x_1',x_2,x_2',\mu_0) \rightarrow  \sigma_{eff}=
{81 \over 64 \int f^2_{uu}({\bf {k_\perp}})
{d {\bf {k_\perp}} \over (2 \pi)^2}}
\simeq 10.9 \, mb\,,
\label{pred}
\end{eqnarray}

a value which turns out to be independent on the momentum scale
$Q$ and on the longitudinal momentum fractions $x_i, x_i'$.
It is remarkable that it compares reasonably well with the 
sets of data shown in Fig.1.  Of course the validity of our simplified result 
is restricted to the valence region where the model is predictive and the
numerical estimate is connected to the ability of the model to capture
(in its wave function) the correct average distribution of the valence quarks 
in transverse space.

\end{itemize}

The $x$-dependence we are discussing does not emerge from the present data, 
probably not accurate enough.  Our study points out therefore to an 
experimental scenario
where more precise measurements in narrow $x_i$ regions could shed
new light on the structure of the proton and on the nature of hard
proton-proton collisions. 
If the $x$-dependence is seen, one will gain, through $\sigma_{eff}$,
a first indication of double parton correlations and a fresh look at the 3D proton structure.



\section{Conclusions}

We have calculated the effective cross section  $\sigma_{eff}$
within a relativistic Poincar\'e covariant quark model.
Extracted from proton-proton scattering data by several
experimental collaborations in the last 30 years, $\sigma_{eff}$
represents a tool to understand double parton scattering
in a p-p collision. Our investigation predicts
a behavior of $\sigma_{eff}$ which, when averaged over the longitudinal 
momentum fractions $x_i$, is consistent with the 
present experimental scenario, in particular with the
sets of data which include the valence region.
However, at the same time, an $x_i$ dependence of
$\sigma_{eff}$ is found, a feature not 
easily read in the available data. We conclude that
the measurement of $\sigma_{eff}$ in restricted $x_i$ ranges
woud lead to a first indication of 
double parton correlations in the proton, addressing a novel
and interesting
aspect of the 3D structure of the nucleon.
The analysis of peculiar
processes where these effects could be most easily seen,
as well as the extension of the model to obtain a better
description of the low-$x$ region, presently studied at LHC,
are in progress. 

\section*{Acknowledgements}

This work was supported in part by 
by the Mineco under contract FPA2013-47443-C2-1-P, by GVA-Prometeo/11/2014/066,
and by CPAN(CSD- 00042). S.S. thanks the Department of Theoretical Physics of 
the University of Valencia for warm hospitality and support. 
M.T. and V.V. thank the INFN, sezione di Perugia and the Department 
of Physics and Geology of the University of Perugia for 
warm hospitality and support.


%
\end{document}